\documentclass[12pt,prd,tightenlines,nofootinbib]{revtex4}
\usepackage{bm}
\usepackage{graphicx}
\usepackage{rotating}
\usepackage{epsfig}
\begin{document}

\title{Triply heavy baryon spectroscopy in the
  relativistic quark model } 

\author{R. N. Faustov}
\author{V. O. Galkin}
\affiliation{
  Federal Research Center ``Computer Science and Control'', Russian Academy of Sciences, 
  Vavilov Street 40, 119333 Moscow, Russia }

\begin{abstract}
Triply heavy baryons are investigated in the framework of
the relativistic quark model based on the quark-diquark picture in the
quasipotential approach in QCD. Masses of the ground and
excited states of  the $\Omega_{ccc}$, $\Omega_{bbb}$, $\Omega_{ccb}$
and $\Omega_{cbb}$ baryons are calculated. Orbital and radial excitations between
the diquark and quark as well as between quarks inside the diquark are
considered. The diquark  internal structure is consistently taken into
account by the form factor of the diquark-gluon interaction expressed
through the overlap integral of the diquark wave functions. The detailed
comparison with previous calculations is given.   
\end{abstract}

\maketitle

\section{Introduction}
\label{sec:intr}

Recently significant experimental progress has been achieved in
studying hadrons with heavy quarks. Many new states of heavy mesons
as well as of heavy baryons 
were observed, some of which have properties implying their exotic
nature (for recent reviews see, e.g., \cite{lcclz,behnstvy,fgs} and
references therein).  A special interest represent the  long-awaited
discoveries very recently made by
the LHCb Collaboration: of the doubly charmed baryon $\Xi_{cc}^{++}$
\cite{lhcbdhb};   of the tetraquark $T_{cc\bar c\bar c}$, composed of
two charm quarks and two charm antiquarks, X(6900) \cite{lhcbcccc}; and
of the doubly charmed tetraquark, $T_{cc}^+$, with a quark content
$cc\bar u\bar d$ \cite{lhcbtcc}. All these new states require the
production of at least two charm quark-antiquark pairs. The next important
step forward will be the discovery of the triply heavy baryons, composed only from
heavy charm and/or bottom quarks, and, thus, requiring the production
of three heavy quark-antiquark pairs. The first observation of the
simultaneous production of three $J/\psi$ mesons in proton-proton
collisions was very recently presented by the CMS Collaboration \cite{cms}. Estimates of the production
cross-section of triply heavy baryons in proton-proton  \cite{prod}
and heavy ion \cite{prod2} collisions  indicate that triply
charmed $\Omega_{ccc}$ baryons  have good chances to be observed  at
LHC.

In this paper we apply the relativistic quark model based on the
quasipotential approach in QCD to calculate the mass spectra of triply
heavy baryons. These baryons contain only heavy quarks and in the literature they are
usually treated as nonrelativistic systems. However, the
investigation of the heavy quark dynamics in heavy 
quarkonia shows that heavy quarks should be treated
relativistically \cite{mass}. Indeed, estimates of the charm quark
velocity $v$ in
charmonia show that it is about one half of the velocity of light $c$,
while the bottom quark velocity in bottomonia is about one third of $c$. Our previous
investigations of meson  \cite{mass,lregge}, baryon
\cite{dhbar,hbar,hbarRegge,sbar} and tetraquark \cite{fgs,htetr} properties showed that relativistic
effects play a very important role. Thus we treat triply heavy baryons
completely relativistically without application of the expansion in
heavy quark velocity. To achieve this goal we use the relativistic
quark-diquark model which was previously developed and applied for the consideration
of heavy \cite{hbar,hbarRegge}, doubly heavy \cite{dhbar} and strange \cite{sbar} baryons. Constructing the triply
heavy baryon we assume that two quarks of the same flavor form a
doubly heavy diquark and the baryon is a relativistic bound system of
this doubly heavy diquark and heavy quark. The masses and wave
functions of diquarks are obtained by solving the relativistic
quasipotential equation with the quark-quark interaction, which is one
half of the quark-antiquark interaction in mesons. The diquark is
considered to be composite, not a point-like object. The diquark  internal structure
is taken into account by calculating the diquark form factor, which
enters the diquark-gluon interaction, and is expressed as the overlap
integral of the diquark wave functions. The account of the diquark
size softens the diquark-gluon interaction thus increasing the baryon
mass. This effect allowed us to get the correct prediction for the
doubly charmed baryon $\Xi_{cc}^{++}$ mass long before its discovery
\cite{dhbar}. It is important to point out that consistent treatment
of the relativistic quark dynamics permitted us to get predictions for
meson \cite{mass,lregge}, baryon  \cite{dhbar,hbar,hbarRegge,sbar} and
tetraquark \cite{fgs} masses and decays in good agreement with
experimental data using the universal set
of model parameters, which we keep fixed in the present calculations
of the triply baryon spectroscopy. Note that in most 
quark models the parameters for description of meson and baryon
properties are varied. 

The paper is organized as follows. In Sec.~\ref{sec:rqdm} we briefly
describe our relativistic quark-diquark model. The quasipotential
equation and quark-quark and quark-diquark interaction potentials are
given. Doubly heavy diquarks are considered in
Sec.~\ref{sec:dhd}. Their masses are calculated up to the first radial
excitation and second orbital excitation. The form factors entering the
diquark-gluon interaction are evaluated and their appropriate
parametrization is given. In Sec.~\ref {sec:thb} we calculate the
masses of the ground, orbitally and radially excited states of triply heavy baryons and
compare our results with previous calculations. Finally,
Sec.~\ref{sec:concl} contains our conclusions. 

\section{Relativistic quark-diquark model}
\label{sec:rqdm}

The relativistic quark-diquark model for description of doubly heavy,
heavy baryon and hyperon spectroscopy based on the quasipotential approach and
quark-diquark picture of baryons was developed and previously used in Refs.
\cite{dhbar,hbar,hbarRegge,sbar}. Here we apply this model for the consideration
of the triply heavy baryon spectroscopy. We use the same assumptions
and model parameters. For completeness we give its
brief outline.  In this approach the complicated relativistic three
body problem is reduced to the solution of two more simple
relativistic two body problems. First, we introduce diquarks which are
considered to be bound states of two quarks. It is assumed
that quarks of the same flavor form a diquark. Second, the
baryon is considered to be a bound system of a diquark and quark.
In quasipotential approach interactions of two quarks in a diquark and of the quark and diquark in a baryon are described by the
diquark wave function $\Psi_{d}$ and by the baryon wave function
$\Psi_{B}$,  respectively.   These wave functions satisfy the
quasipotential equation of the Schr\"odinger type \cite{mass}
\begin{equation}
\label{quas}
{\left(\frac{b^2(M)}{2\mu_{R}}-\frac{{\bf
p}^2}{2\mu_{R}}\right)\Psi_{d,B}({\bf p})} =\int\frac{d^3 q}{(2\pi)^3}
 V({\bf p,q};M)\Psi_{d,B}({\bf q}),
\end{equation}
with the relativistic reduced mass given by
\begin{equation}
\mu_{R}=\frac{M^4-(m^2_1-m^2_2)^2}{4M^3},
\end{equation}
and the center-of-mass system relative momentum squared on mass shell 
defined by
\begin{equation}
{b^2(M) }
=\frac{[M^2-(m_1+m_2)^2][M^2-(m_1-m_2)^2]}{4M^2},
\end{equation}
where $M$ is the bound state mass (diquark or baryon),
$m_{1,2}$ are the masses of  quarks ($q_1$ and $q_2$) which form
the diquark or of the  diquark ($d$) and  quark ($q$) which form
the baryon ($B$), and ${\bf p}$  is their relative momentum. 

To construct the kernel 
$V({\bf p,q};M)$ in Eq.~(\ref{quas}), which is the quasipotential operator of
the quark-quark or quark-diquark interactions, it is assumed that the effective
interaction is the sum of the usual one-gluon exchange term and the mixture
of long-range vector and scalar linear confining potentials \cite{hbar,hbarRegge}. The
vector confining potentials contain additional effective Pauli terms, which
introduce anomalous chromomagnetic moments of quarks and diquarks. 

The quark-quark ($qq$) interaction quasipotential for the diquark is
given by
 \begin{equation}
\label{qpot}
V({\bf p,q};M)=\bar{u}_{1}(p)\bar{u}_{2}(-p){\cal V}({\bf p}, {\bf
q};M)u_{1}(q)u_{2}(-q),
\end{equation}
with
\[
{\cal V}({\bf p,q};M)=\frac12\left[\frac43\alpha_sD_{ \mu\nu}({\bf
k})\gamma_1^{\mu}\gamma_2^{\nu}+ V^V_{\rm conf}({\bf k})
\Gamma_1^{\mu}({\bf k})\Gamma_{2;\mu}(-{\bf k})+
 V^S_{\rm conf}({\bf k})\right],
\]
where $\alpha_s$ is the QCD coupling constant, $D_{\mu\nu}$ is the  
gluon propagator in the Coulomb gauge, ${\bf k=p-q}$; $\gamma_{\mu}$
and $u(p)$ are  the Dirac matrices and spinors, and the effective long-range vector vertex of the quark is
defined by \cite{mass}
\begin{equation}
\Gamma_{\mu}({\bf k})=\gamma_{\mu}+
\frac{i\kappa}{2m}\sigma_{\mu\nu}\tilde k^{\nu}, \qquad \tilde
k=(0,{\bf k}),
\end{equation}
and $\kappa$ is the anomalous chromomagnetic moment of quarks.

The quark-diquark ($qd$) interaction quasipotential in the baryon has
the form
\begin{eqnarray}
\label{dpot}
V({\bf p,q};M)&=&\frac{\langle d(P)|J_{\mu}|d(Q)\rangle}
{2\sqrt{E_d(p)E_d(q)}} \bar{u}_{q}(p)  
\frac43\alpha_sD_{ \mu\nu}({\bf 
k})\gamma^{\nu}u_{q}(q)\cr
&&+\psi^*_d(P)\bar u_q(p)J_{d;\mu}\Gamma_q^\mu({\bf k})
V_{\rm conf}^V({\bf k})u_{q}(q)\psi_d(Q)\cr 
&&+\psi^*_d(P)
\bar{u}_{q}(p)V^S_{\rm conf}({\bf k})u_{q}(q)\psi_d(Q), 
\end{eqnarray}
where  $\langle
d(P)|J_{\mu}|d(Q)\rangle$ is the vertex of the 
diquark-gluon interaction which takes into account the diquark
internal structure, $J_{d;\mu}$ is the effective long-range vector
vertex of the diquark, the diquark momenta are $P=(E_d(p),-{\bf p})$,
$Q=(E_d(q),-{\bf q})$ with $E_d(p)=\sqrt{{\bf
    p}^2+M_d^2}$, and $\psi_d(P)$ is the diquark wave
function \cite{hbar}.

The vector and scalar confining potentials in the nonrelativistic
limit  in configuration space are the linear potentials with   the mixing coefficient $\varepsilon$:
\begin{equation}
V^V_{\rm conf}(r)=(1-\varepsilon)(Ar+B),\qquad
V^S_{\rm conf}(r) =\varepsilon (Ar+B),
\end{equation}
and their sum is just the
usual static Cornell-like potential 
\begin{equation}
V(r)=-\frac43\frac{\alpha_s}{r}+Ar+B,
\end{equation}
with the freezing QCD coupling constant 
\begin{equation}
  \label{eq:alpha}
  \alpha_s(\mu^2)=\frac{4\pi}{\displaystyle\left(11-\frac23n_f\right)
\ln\frac{\mu^2+M_B^2}{\Lambda^2}}.
\end{equation}
The scale $\mu=2\mu_{NR}$ is chosen to be twice the nonrelativistic reduced mass
$\mu_{NR}=m_1m_2/(m_1+m_2)$, the background mass is taken $M_B=0.95$~GeV, and the parameter 
$\Lambda=413$~MeV  was fixed from the light meson spectroscopy \cite{lregge}.

All parameters of the model  are kept
fixed from previous  calculations of meson and baryon properties
\cite{mass,dhbar,hbar}. The constituent heavy quark masses are $m_c=1.55$
GeV, $m_b=4.88$ GeV  and the parameters of the linear potential are
$A=0.18$ GeV$^2$ and $B=-0.3$ GeV,  the value of the mixing coefficient of vector and scalar
confining potentials is $\varepsilon=-1$ and the
anomalous chromomagnetic quark moment is $\kappa=-1$.  Note that the long-range chromomagnetic
contribution to the potential, which is proportional to $(1+\kappa)$,
vanishes for the chosen value of $\kappa=-1$.

\section{Doubly heavy diquarks}
\label{sec:dhd}

At first step we consider the doubly heavy diquarks and  calculate
their masses and form factors. We assume that in triply heavy baryons two quarks of the
same flavor form a doubly heavy diquark. We solve the quasipotential
equation (\ref{quas}) numerically with the complete relativistic
potential (\ref{qpot}). Since a diquark is composed of heavy quarks of the same
flavor it is necessary to take into account the Pauli principle. The
total baryon wave function must be antisymmetric. It is antisymmetric in color, thus the rest of the wave function must be
symmetric. The symmetric form of the flavor part of the considered
doubly heavy diquark implies that the product of the spin and orbital
parts is also symmetric. For the $S$ and $D$ states, which
have orbitally symmetric wave functions, the
diquark spin wave function must be also symmetric and, thus, the
diquark spin is 1. For the $P$ states with antisymmetric orbital wave
function the diquark spin wave function must be also antisymmetric and
the diquark spin is 0. The resulting value of the total momentum of
the diquark is $j= 1$ for
the $S$ and $P$ states, while values $j=1,2,3$ are possible for  the $D$ states.
Note that we do not consider higher orbital excitations of the quarks
inside the diquark.

\begin{table}
  \caption{Masses $M$ and form factor  parameters of the doubly heavy
   diquarks. }
  \label{tab:dhdmass}
\begin{ruledtabular}
\begin{tabular}{ccccc}
Quark&State&   $M$ &$\xi$ & $\zeta$
 \\
content&$nl_j$ & (MeV)& (GeV)& (GeV$^2$)  \\
\hline
$cc$&$1s_1$ & 3226 & 1.30 & 0.42  \\
&$1p_1$ & 3460 &0.74 & 0.315  \\
&$2s_1$ & 3535 &0.67 & 0.19  \\
     &$1d_{1,2,3}$ & 3704 &0.39 & 0.42  \\
  &$2p_1$ & 3712 &0.60 & 0.155  \\
\hline
$bb$ &$1s_1$ & 9778 & 1.30 & 1.60\\
 &$1p_1$ & 9944 & 0.90 & 0.59\\
     &$2s_1$ & 10015 & 0.85 & 0.31\\
  &$1d_{1,2,3}$ & 10123 &0.49 & 0.59  \\
  &$2p_1$ & 10132 &0.65 & 0.215  \\
  \end{tabular}
\end{ruledtabular}
\end{table}

The calculated
masses of the ground and excited states of diquarks are presented in
Table~\ref{tab:dhdmass}. We also give the values of
the parameters $\xi$ and $\zeta$. They  parameterize with high accuracy the
$r$-dependence of the form factor $F(r)$ in the vertex of the  diquark-gluon interaction
(\ref{dpot}), which is calculated through the overlap integral of the
diquark wave functions \cite{hbar}. It is expressed by
\begin{equation}
  \label{eq:fr}
  F(r)=1-e^{-\xi r -\zeta r^2},
\end{equation}
and takes the internal structure of a diquark into account \cite{hbar}
smearing the diquark-gluon interaction.   In this Table we use the
lowercase letters to denote diquark quantum numbers. This is done to
distinguish them form the quark-diquark excitations for which we
reserve the
uppercase letters. Here $n=n_r+1$, where $n_r$ is
the radial quantum number (the number of nodes of the wave function); $l=s,p,d\dots$ is the orbital momentum and
$j$ is the total momentum of the diquark. The calculations show that
the masses of the $d$ states with $j=1,2,3$ differ by less than 1 MeV,
thus we consider their masses to be equal and give only one value.

\section{Triply heavy baryons}
\label{sec:thb}

At the second step we calculate the masses of the triply heavy baryons as the bound
states of a heavy quark and doubly heavy diquark.  Evaluating
the baryon masses we treat all relativistic contributions
nonperturbatively. The quark-diquark
quasipotential contains the relativistic contributions both to the
spin-independent  $V_{\rm SI}$ and spin-dependent $V_{\rm SD}$ parts
\begin{equation}
  \label{eq:vp}
  V(r)= V_{\rm SI}(r)+ V_{\rm SD}(r).
\end{equation}
The spin-independent part determines the position of centers of
gravity of the baryon levels, while the spin-dependent part is
responsible for their fine and hyperfine splittings. 
These parts
are expressed trough the static potential and its derivatives. The explicit
expressions for these potentials  can be found in Refs.~
\cite{hbar,hbarRegge}. It is important to point out that, as it was
already noted in the previous section, the diquark form factor $F(r)$ smears the
diquark-gluon interaction, thus, accounting for its internal structure.
As a result, in the nonrelativistic limit  the one-gluon exchange part of the
quark-diquark potential is modified  and has the form of the smeared
Coulomb-like potential
\begin{equation}
  \label{eq:mcp}
  \hat V_{\rm Coul}(r)=-\frac43\alpha_s\frac{ F(r)}{r},
\end{equation} 
with $F(r)$ given by Eq.~(\ref{eq:fr}).

The spin-dependent part of the quasipotential contains  the
spin-orbit, tensor and spin-spin interactions. It has the following form \cite{hbarRegge}
\begin{equation}
  \label{eq:vsd}
   V_{\rm SD}(r)=a_1\, {\bf L}{\bf S}_d+a_2\, {\bf L}{\bf S}_Q+
b \left[-{\bf S}_d{\bf S}_Q+\frac3{r^2}({\bf S}_d{\bf r})({\bf
    S}_Q{\bf r})\right]+ c\, {\bf S}_d{\bf S}_Q,
\end{equation}
where $ {\bf L}$ is the orbital angular momentum; ${\bf S}_d$ and
${\bf S}_Q$ are the diquark and quark spin operators,
respectively. 
The coefficients $a_1$, $a_2$, $b$ and $c$ are expressed
through the corresponding derivatives of the smeared Coulomb and
confining potentials \cite{hbarRegge}. The smearing of the one-gluon
exchange potential (\ref{eq:mcp}) naturally softens singularities in the relativistic
quasipotential in configuration space and allows us to solve numerically
the quasipotential equation in its complete relativistic form.
Note that both the one-gluon exchange and confining potentials contribute
to the quark-diquark spin-orbit interaction. The presence of the spin-orbit
${\bf L}{\bf S}_Q$ and of  tensor terms in the quark-diquark potential  leads to a mixing of states with
the same total angular momentum $J$ and parity $P$ but different
values of the
diquark  angular momentum  (${\bf L}+{\bf S}_d$). We consider
such mixing  in the same way as in the case of doubly heavy baryons \cite{dhbar}.

We calculate  masses of all triply heavy  baryons:
$\Omega_{ccc}$, $\Omega_{bbb}$, $\Omega_{ccb}$ and $\Omega_{cbb}$. Their calculated spectra  are given in
Tables~\ref{omegaccc}--\ref{omegacbb}.  In the left hand half of these
tables we give states with the positive parity and in the right one with
the negative parity. We use
the standard notations for the baryon
states $J^P$, where
$J$ and $P$ are the baryon total spin and parity, respectively.  The
composition of the baryon state is given by $NLnl_j$, where the capital
letters denote quantum numbers of the quark-diquark system and the
lowercase letters the diquark state. $N$ or $n$ is the radial quantum
number (the number of nodes of the wave function) plus one and $L$ or $l$ is the orbital quantum number. 

For the baryons composed
of identical quarks ($\Omega_{ccc}$ and $\Omega_{bbb}$) there is an
additional complication. It is
necessary  to take into account the Pauli principle not only for the
diquark but also for the entire baryon. It requires the
total wave function to be antisymmetric.  The color wave function of
the baryon is antisymmetric. The flavor part is symmetric. This means that the spin-momentum part
of the wave function must be fully symmetric. For the ground $1S1s$ state, this part is symmetric
in momentum and, thus, the spin wave functions must be also fully
symmetric, which corresponds to the total baryon spin $3/2$.
Therefore, only the  $3/2^+$ ground state is possible. The lightest $1/2^+$
state should contain excitations and has a significantly larger mass. The excited states are
combinations of  orbital and/or radial excitations of the diquark
and/or quark-diquark bound systems with the  fully symmetric ($3/2$) or
 mixed symmetry ($1/2$) spin wave functions. Symmetric combinations
 such as, e.g., $|2S1s\rangle_+=(|2S1s\rangle+ |1S2s\rangle)/\sqrt2$,  
$|1D1s\rangle_+=(|1D1s\rangle+ |1S1d\rangle)/\sqrt2$,  $|2S2s\rangle$ are combined
with fully symmetric spin $3/2$ wave function. While antisymmetric
combinations such as, e.g., $|2S1s\rangle_-=(|2S1s\rangle- |1S2s\rangle)/\sqrt2$,  
$|1D1s\rangle_-=(|1D1s\rangle- |1S1d\rangle)/\sqrt2$,
$|1P2s\rangle_-=(|1P2s\rangle- |1S2p\rangle)/\sqrt2$, are combined
with the mixed symmetry spin $1/2$ wave functions. The details can be
found in Ref.~\cite{sgmg}.

\begin{table}
\caption{Masses of the $\Omega_{ccc}$ states (in MeV).}
\label{omegaccc}
\begin{ruledtabular}
\begin{tabular}{ccc@{\ \ \ \ \ \ \ \ \ \ }||ccc}
$J^P$& $NLnl$& Mass&$J^P$& $NLnl$& Mass\\
\hline
  $\frac12^+$& $2S1s_-,1P1p$& 5230&$\frac12^-$& $1S1p,1P1s$& 5010\\
     &$1D1s$& 5278&&$1P2s_-,2S1p_-$&5370\\
  $\frac32^+$& $1S1s$& 4712&&$2S1p_+$&5385\\
     &$1S2s_+$&5137&&$1D1p_+$&5520\\
     &$1D1s_+$&5267& $\frac32^-$&$1S1p,1P1s$& 5029\\
     &$1D1s_-,1P1p$&5277&&$1P2s_-,2S1p_-$&5379\\
 &$2S2s$&5541&&$2S1p_+$&5394\\
   $\frac52^+$&$1D1s_-,1P1p$&5278&   &$1D1p_+$&5522\\
     &$1D1s_+$&5290& $\frac52^-$&$1F1s$&5519\\
  $\frac72^+$&$1D1s_+$&5291&&$1P1d_+$&5523\\
     &&& $\frac72^-$&$1F1s$&5517\\
  &&&&$1P1d_+$&5526\\ 
\end{tabular}
\end{ruledtabular}
\end{table}

\begin{table}
\caption{Masses of the $\Omega_{bbb}$ states (in MeV).}
\label{omegabbb}
\begin{ruledtabular}
\begin{tabular}{ccc@{\ \ \ \ \ \ \ \ \ \ }||ccc}
$J^P$& $NLnl$& Mass&$J^P$& $NLnl$& Mass\\
\hline
  $\frac12^+$& $2S1s_-,1P1p$& 14877&$\frac12^-$& $1S1p,1P1s$& 14698\\
     &$1D1s$& 14912&&$1P2s_-,2S1p_-$&14991\\
  $\frac32^+$& $1S1s$& 14468&&$2S1p_+$&15042\\
     &$1S2s_+$&14815&&$1D1p_+$&15088\\
     &$1D1s_-,1P1p$ &14893& $\frac32^-$&$1S1p,1P1s$& 14702\\
   &$1D1s_+$ &14905&&$1P2s_-,2S1p_-$&14922\\
 &$2S2s$&15123&&$2S1p_+$&15031\\
   $\frac52^+$&$1D1s_-,1P1p$&14895&   &$1D1p_+$&15089\\
     &$1D1s_+$&14907& $\frac52^-$&$1F1s$&15081\\
  $\frac72^+$&$1D1s_+$&14909&&$1P1d_+$&15086\\
     &&& $\frac72^-$&$1F1s$&15082\\
  &&&&$1P1d_+$&15089\\
\end{tabular}
\end{ruledtabular}
\end{table}

\begin{table}
\caption{Masses of the $\Omega_{ccb}$ states (in MeV).}
\label{omegaccb}
\begin{ruledtabular}
\begin{tabular}{ccc@{\ \ \ \ \ \ \ \ \ \ }||ccc}
$J^P$& $NLnl$& Mass&$J^P$& $NLnl$& Mass\\
\hline
$\frac12^+$& $1S1s$& 7984&$\frac12^-$& $1P1s$& 8250\\
     &$1S2s$&8361&&$1S1p$&8266\\
     &$2S1s$&8405&&$1P1s$&8268\\
    &$1D1s$&8472&&$1S2p$&8550\\
     &$1P1p$&8505&&$2P1s$&8583\\
     &$1P1p$&8511&&$2P1s$&8591\\
     &$1S1d$&8531&&$1P2s$&8592\\
  $\frac32^+$& $1S1s$&7999&&$1P2s$&8595\\
     &$1S2s$&8366&$\frac32^-$& $1P1s$&8262\\
     &$2S1s$&8412&&$1P1s$&8268\\
     &$1D1s$&8474&&$1S1p$&8273\\
     &$1D1s$&8476&&$1S2p$&8554\\
     &$1P1p$&8506&&$2P1s$&8587\\
     &$1P1p$&8510&&$2P1s$&8591\\
     &$1S1d$&8534&&$1P2s$&8591\\
  $\frac52^+$& $1D1s$&8473&&$1P2s$&8594\\
     & $1D1s$&8476& $\frac52^-$&$1P1s$&8267\\
     &$1P1p$&8508&&$2P1s$&8590\\
     &$1S1d$&8536&&$1P2s$&8592\\
 $\frac72^+$& $1D1s$& 8473&$\frac72^-$& $1F1s$& 8647\\ 
  &$1S1d$&8538&&\\
\end{tabular}
\end{ruledtabular}
\end{table}

\begin{table}
\caption{Masses of the $\Omega_{cbb}$ states (in MeV).}
\label{omegacbb}
\begin{ruledtabular}
\begin{tabular}{ccc@{\ \ \ \ \ \ \ \ \ \ }||ccc}
$J^P$& $NLnl$& Mass&$J^P$& $NLnl$& Mass\\
\hline
$\frac12^+$& $1S1s$& 11198&$\frac12^-$& $1P1s$& 11414\\
     &$1S2s$&11507&&$1S1p$&11506\\
     &$1S1d$&11622&&$1P1s$&11540\\
    &$2S1s$&11690&&$1S2p$&11654\\
    &$1P1p$&11692&&$1P2s$&11778\\
      &$1P1p$&11714&&$1P2s$&11796\\
     &$1D1s$&11796&&$2S1p$&11893\\
  $\frac32^+$& $1S1s$&11217&$\frac32^-$& $1S1p$&11424\\
     &$1S2s$&11515&&$1P1s$&11535\\
     &$1S1d$&11629&&$1P1s$&11541\\
     &$2S1s$&11700&&$1S2p$&11660\\
     &$1P1p$&11707&&$1P2s$&11788\\
     &$1P1p$&11717&&$1P2s$&11795\\
     &$1D1s$&11797&&$2S1p$&11897\\
     &$1D1s$&11807&$\frac52^-$&$1P1s$&11543\\
  $\frac52^+$& $1S1d$&11632&&$1P2s$&11795\\
     & $1P1p$&11715& $\frac72^-$&$1P1d$&11903\\
     &$1D1s$&11806&\\
     &$1D1s$&11807&\\
 $\frac72^+$& $1S1d$& 11635&\\
\end{tabular}
\end{ruledtabular}
\end{table}

Masses of the ground states of $\Omega_{ccc}$, $\Omega_{bbb}$,
$\Omega_{ccb}$ and $\Omega_{cbb}$ baryons were calculated in many
papers based on different approaches \cite{pemp,ypos,sgmg,sr,rp,llz,qrs,
m,sr1,mpm,bdmo,sr2,ceot,pacs,blb,adjkk,vvg,sb,hhkr,mart,zh,aas,w,mmmp,rbm,tmv,yckrsx,j,wcg,wclwg}. The
predicted masses of the ground state $3/2^+$ $\Omega_{ccc}$ baryons
range from 4670 to 4990 MeV and masses of the $3/2^+$ $\Omega_{bbb}$
baryons range from 13280 to 14834 MeV.  Our predictions for these
masses: 4712 MeV and 14468 MeV, respectively, are well inside both ranges.

Excited states received significantly less
attention. In Tables~\ref{occccomp}-\ref{ocbbcomp} we compare our
predictions with previous calculations \cite{pemp,ypos,sgmg,sr,rp,llz,qrs,
m,mpm,bdmo,sr1,sr2} for the masses of $\Omega_{ccc}$, $\Omega_{bbb}$,
$\Omega_{ccb}$ and $\Omega_{cbb}$ baryons. Masses of the triply heavy
baryons were calculated using lattice QCD with dynamical light quark
fields in Refs.~\cite{pemp,m,mpm,bdmo}. Our predictions for the masses
of the ground  and excited states of the $\Omega_{ccc}$ baryon are
lower than lattice  \cite{pemp} results by about 50--150 MeV, however
the structure of our excited spectrum is close to the lattice one. For
the $\Omega_{bbb}$ baryon the agreement of our predictions with
lattice results \cite{m} is even better. Masses of only ground states
$1/2^+$ and $3/2^+$  of  $\Omega_{ccb}$ and $\Omega_{cbb}$ baryons
were calculated on the lattice \cite{mpm,bdmo}. They agree well with
our results. The constituent quark model, which employs the  Gaussian
expansion method and the variational principle to solve the  nonrelativistic three-body
problem, was used in Ref.~\cite{ypos}  to compute the mass spectra of
triply heavy baryons. The renormalization group procedure for
effective particles was applied for studying baryons with heavy quarks
in Ref.~\cite{sgmg}. The hypercentral constituent quark model was
employed in Refs.~\cite{sr,sr1,sr2}. For the calculation of the triply heavy baryon masses Refs.~\cite{llz,rp} used the
nonrelativistic quark model with the harmonic oscillator wave
functions and perturbative account of the relativistic
corrections. The relativistic Faddeev equation with the rainbow-ladder
truncated kernel was employed in Ref.~\cite{qrs}.

\begin{table}
\caption{Comparison with previous theoretical predictions for the masses of the $\Omega_{ccc}$ states (in MeV).}
\label{occccomp}
\begin{ruledtabular}
\begin{tabular}{ccccccccc}
$J^P$&Our & \cite{pemp}&\cite{ypos}&\cite{sgmg}&\cite{sr}&\cite{rp}&\cite{llz}&\cite{qrs}\\
\hline
  $\frac12^+$& 5230&5397(13)&5376&5358&5473&5325&5352\\
     &5278&5403(14)&&&&5332&5373\\
  $\frac32^+$& 4712&4761(6)&4798&4797&4806&4965&4828&4760\\
     &5137&5315(31)&5286&5309&5300&5313&5285&5150\\
     &5267&5428(13)&5376&5358&5448&&5368\\
     &5277&5463(13)&&&&&5412\\
  $\frac52^+$& 5278&5404(15)&5376&5358&5416&5329&5392\\
     &5290&5462(15)&&&&5343&5433\\
  $\frac72^+$& 5291&5395(49)&5376&5358&5375&5331&5418\\
  $\frac12^-$& 5010&5118(9)&5129&5103&5012&5155&5142\\
     &5370&5610(31)&5525&&5607\\
     &5385&5629(43)\\
  $\frac32^-$& 5029&5122(13)&5129&5103&4991&5160&5162&5027\\
     &5379&5660(31)&5525&&5584\\
     &5394&5722(44)\\
  $\frac52^-$& 5519&5514(64)&5558&&4965&&\\
     &5523&5707(25)&&&5584\\
  $\frac72^-$& 5517&5679(28)&&&5829&&\\
\end{tabular}
\end{ruledtabular}
\end{table}

\begin{table}
\caption{Comparison with previous theoretical predictions for the masses of the $\Omega_{bbb}$ states (in MeV).}
\label{obbbcomp}
\begin{ruledtabular}
\begin{tabular}{ccccccccc}
$J^P$&Our & \cite{m}&\cite{ypos}&\cite{sgmg}&\cite{sr}&\cite{rp}&\cite{llz}&\cite{qrs}\\
\hline
  $\frac12^+$& 14877&14938(18)&14894&14896&15306&15097&14971\\
     &14912&14953(17)&&&&15102&14959\\
  $\frac32^+$& 14468&14371(12)&14396&14347&14496&14834&14432&14370\\
     &14815&14840(14)&14805&14832&15154&15089&14848&14980\\
     &14893&14958(18)&14894&14896&15300&&14975\\
     &14905&15005(20)&&&&&15016\\
  $\frac52^+$& 14895&14964(18)&14894&14896&15293&15101&14981\\
     &14907&15007(20)&&&&15109&15022\\
  $\frac72^+$& 14909&14969(17)&14894&14896&15286&15101&14988\\
  $\frac12^-$& 14698&14706(9)&14688&14645&14944&14975&14773\\
     &14991&&15016&&\\
  $\frac32^-$& 14702&14714(9)&14688&14645&14937&14976&14779&14771\\
     &14922&&15016&&\\
  $\frac52^-$& 15081&&15038&&14931&&\\
  $\frac72^-$& 15082&&&&15641&&\\
\end{tabular}
\end{ruledtabular}
\end{table}

\begin{table}
\caption{Comparison with previous theoretical predictions for the masses of the $\Omega_{ccb}$ states (in MeV).}
\label{occbcomp}
\begin{ruledtabular}
\begin{tabular}{ccccccccc}
$J^P$&Our & \cite{mpm}&\cite{bdmo}&\cite{ypos}&\cite{sgmg}&\cite{sr1}&\cite{rp}&\cite{qrs}\\
\hline
  $\frac12^+$& 7984&8005(17)&8007(29)&8004&8301&8005&8245&7867\\
     &8361&&&8455&8647&8621&8537&8337\\
   &8405&&&8536&&8848&\\
  $\frac32^+$& 7999&8026(18)&8037(29)&8023&8301&8049&8265&7963\\
     &8366&&&8468&8600&8637&8553&8427\\
     &8412&&&8536&8647&8831&\\
  $\frac52^+$& 8473&&&8536&8647&8808&8568\\
     &8476&&&&&&8571\\
  $\frac72^+$& 8473&&&8538&8647&8780&8568\\
    &8538&&&&&&8653\\
  $\frac12^-$& 8250&&&8306&8491&8400&8418&8164\\
     &8266&&&&&&8422\\
  $\frac32^-$& 8262&&&8306&8491&8383&8420&8275\\
     &8268&&&&&&8422\\
  $\frac52^-$& 8267&&&8311&8491&8365&8432\\
\end{tabular}
\end{ruledtabular}
\end{table}

\begin{table}
\caption{Comparison with previous theoretical predictions for the masses of the $\Omega_{cbb}$ states (in MeV).}
\label{ocbbcomp}
\begin{ruledtabular}
\begin{tabular}{ccccccccc}
$J^P$&Our & \cite{mpm}&\cite{bdmo}&\cite{ypos}&\cite{sgmg}&\cite{sr2}&\cite{rp}&\cite{qrs}\\
\hline
  $\frac12^+$& 11198&11194(17)&11195(28)&11200&11218&11231&11535&11077\\
     &11507&&&11607&11585&11757&11787&11603\\
   &11622&&&11677&11626&11934&\\
  $\frac32^+$& 11217&11211(18)&11229(28)&11221&11218&11296&11554&11167\\
     &11515&&&11622&11585&11779&11798&11703\\
     &11629&&&11677&11626&11928&\\
  $\frac52^+$& 11632&&&11677&11626&11919&11823\\
     &11715&&&&&&11831\\
  $\frac72^+$& 11635&&&11688&11626&11909&11810\\
  $\frac12^-$& 11414&&&11482&11438&11573&11710&11413\\
     &11506&&&&&&11757\\
  $\frac32^-$& 11424&&&11482&11438&11566&11711&11523\\
     &11535&&&&&&11759\\
  $\frac52^-$& 11543&&&11569&11601&11558&11762\\
\end{tabular}
\end{ruledtabular}
\end{table}

\section{Conclusions}
\label{sec:concl}

In this paper we applied the relativistic quark model based on the
quasipotential approach in QCD for the calculation of the mass spectra
of triply heavy baryons. The relativistic quark-diquark approximation
was used to reduce a very complicated relativistic three-body problem
for the subsequent solution of two more simple two-body problems:
first, calculation of the diquark properties and then considering baryon as a
quark-diquark bound system. Such an
approach was previously successfully applied for the calculation of
the masses of the ground and excited states of doubly heavy
\cite{dhbar}, heavy \cite{hbarRegge} and
strange baryons \cite{sbar}. It is important to emphasize that all parameters of
the model were kept fixed from the previous calculations and no new
parameters were introduced.  We assumed
that two identical heavy quarks form a doubly heavy 
diquark. Masses and wave functions of the ground and excited states of
such diquarks were calculated. The internal structure of the diquark
was taken into account by the form factor of the diquark-gluon
interaction which was calculated as the overlap integral of the
diquark wave functions. The internal structure of the diquark  was
found to be very important for obtaining the correct prediction for the mass of
the doubly heavy baryon $\Xi_{cc}$ \cite{dhbar}.  It also allows to
obtain local completely relativistic quark-diquark quasipotential
without fictitious singularities.

We solved numerically the corresponding
quasipotential equation and obtained the 
masses of the ground and excited states of $\Omega_{ccc}$, $\Omega_{bbb}$,
$\Omega_{ccb}$ and $\Omega_{cbb}$ baryons. Excited states
with total spin up to $J=7/2$ both with positive and negative parity
were considered.      The calculated masses were
compared with previous lattice QCD \cite{pemp,m,mpm,bdmo} and quark
model calculations. Reasonable agreement with lattice results was found.

\acknowledgments
The authors are grateful to 
D.~Ebert, J. K\"orner and  A. Martynenko
for  useful discussions.

\end{document}